\documentclass[aps,prd,reprint,nofootinbib,showkeys,showpacs,superscriptaddress,longbibliography,floatfix]{revtex4-1}

\pdfoutput=1

\usepackage{amsmath}
\usepackage{amssymb}
\usepackage[american]{babel}
\usepackage{booktabs}
\usepackage{dcolumn}  
\usepackage[T1]{fontenc}  
\usepackage{graphicx}  
\usepackage[colorlinks]{hyperref}  
\usepackage[utf8]{inputenc}  
\usepackage{multirow}
\usepackage{txfonts}  
\usepackage{url}
\usepackage[dvipsnames]{xcolor}

\newcommand{\afillone}{Coordenação de Astronomia e Astrofísica, Observatório Nacional, 20921-400, Rio de Janeiro -- RJ, Brazil}
\newcommand{\afilltwo}{Instituto de Astronomía, Universidad Nacional Autónoma de México, Box 70-264, México City, México}
\newcommand{\lcdm}{$\Lambda$CDM}
\newcommand{\ltcdm}{$\Lambda(t)$CDM}
\newcommand{\OmegaBo}{\Omega_{\mathrm{b},0}}
\newcommand{\OmegaDMo}{\Omega_{\mathrm{dm},0}}
\newcommand{\OmegaDEo}{\Omega_{\mathrm{de},0}}
\newcommand{\OmegaMo}{\Omega_{\mathrm{m},0}}

\newcommand{\PMN}{\texttt{PyMultiNest}}
\newcommand{\MN}{\textsc{MultiNest}}
\newcommand{\N}[2] {\mathcal{N} \left( #1, #2 \right)}
\newcommand{\U}[2] {\mathcal{U} \left( #1, #2 \right)}

\newcolumntype{d}[1]{D{.}{.}{#1}}
\newcolumntype{v}[1]{D{,}{,\ }{#1}}

\hypersetup{
    citecolor = MidnightBlue,
    linkcolor = MidnightBlue,
    urlcolor = MidnightBlue
}

\begin{document}

\title{Bayesian comparison of nonstandard cosmologies using type Ia supernovae and BAO data}
\author{B. Santos}
\email{\tt thoven@on.br}
\affiliation{\afillone}
\author{N. Chandrachani Devi}
\email{\tt chandrachaniningombam@astro.unam.mx}
\affiliation{\afilltwo}
\affiliation{\afillone}
\author{J. S. Alcaniz}
\email{\tt alcaniz@on.br}
\affiliation{\afillone}

\begin{abstract}
    We use the most recent type Ia supernovae (SNe Ia) observations to perform a statistical comparison between the standard {\lcdm} model and its extensions [$w$CDM and $w(z)$CDM] and some alternative cosmologies: namely, the Dvali--Gabadadze--Porrati (DGP) model, a power-law $f(R)$ scenario in the metric formalism and an example of vacuum decay [\ltcdm] cosmology in which the dilution of pressureless matter is attenuated with respect to the usual $a^{-3}$ scaling due to the interaction of the dark matter and dark energy fields.
    We perform a Bayesian model selection analysis using the {\MN} algorithm.
    To obtain the posterior distribution for the parameters of each model, we use the joint light-curve analysis (JLA) SNe Ia compilation containing 740 events in the interval $0.01 < z < 1.3$ along with current measurements of baryon acoustic oscillations (BAO).
    The JLA data are analyzed with the SALT2 light-curve fitter and the model selection is then performed by computing the Bayesian evidence of each model and the Bayes factor of the {\lcdm} cosmology related to the other models.
    The results indicate that the JLA data alone are unable to distinguish the standard {\lcdm} model from some of its alternatives but its combination with current measurements of baryon acoustic oscillations shows up an ability to distinguish them.
    In particular, the DGP model is practically not supported by both the BAO and the joint JLA + BAO data sets compared to the standard scenario.
    Finally, we provide a rank order for the models considered.
\end{abstract}

\keywords{
    statistics: model selection
    statistics: Bayesian inference
    statistics: parameter estimation
    cosmology: observations -- distance scale -- cosmological parameters;
    cosmology: theory -- dark energy;
    cosmology: theory -- non-standard cosmology;
    cosmology: curvature;
}

\date{\today}

\maketitle

\section{Introduction}

Almost two decades ago, distance measurements of type Ia supernovae (SNe Ia) provided the first direct evidence for a late-time cosmic acceleration~\citep{Riess1998, Perlmutter1999}.
Nowadays, this phenomenon is also confirmed from independent data, such as, for instance, the most recent measurements of the baryon acoustic oscillations (BAO) from galaxy surveys (see, e.g., \cite{Aubourg2015} for a recent review on BAO measurements).
From the theoretical side, however, the absence of a firm physical mechanism responsible for the present acceleration of the Universe has given rise to a number of alternative explanations.

In general, mechanisms of cosmic acceleration are explored in two different ways: either introducing a new field in the framework of the Einstein's general theory of relativity (GR), the dark energy, or introducing modifications in GR at very large scales.
In the general relativistic framework, the simplest explanation is to posit the existence of a cosmological constant $\Lambda$, a spatially homogeneous component whose pressure and energy density are related by $p_\Lambda = w\rho_\Lambda$, with the equation of state (EoS) parameter $w = -1$.
However, as is well known, the standard {\lcdm} model (cosmological constant $\Lambda$ plus cold dark matter) provides a good fit for a large number of observational data sets without addressing some important theoretical issues, such as the fine-tuning of the $\Lambda$ value and the cosmic coincidence problems~\citep{Weinberg1989, Sahni2000, Padmanabhan2003}.
If the cosmological term $\Lambda$ is null or it is not decaying in the course of the expansion, as discussed in the vacuum decay or $\Lambda(t)$ cosmologies~\citep{Ozer1986, Freese1987, Carvalho1992, Alcaniz2005, Carneiro2006}, an alternative possibility (which also does not address the above issues) is to assume the presence of an extra degree of freedom in the form of a minimally coupled scalar field $\phi$ (quintessence field).
Among other things, what observationally may distinguish $\Lambda$ or $\Lambda(t)$ from $\phi$ is the time dependency of the EoS parameter of quintessence fields, whose behavior has been parametrized phenomenologically by several authors (see, e.g., \cite{Efstathiou1999, Chevallier2001, Linder2003, Barboza2008, Barboza2009} and references therein).

The observed cosmic acceleration can also be seen as the first evidence of a breakdown of GR on large scales rather than a manifestation of another ingredient in the cosmic budget.
The most usual examples of cosmologies derived from modified or extended theories of gravity include $f(R)$ models, in which terms proportional to powers of the Ricci scalar $R$ are added to the Einstein-Hilbert Lagrangian~\citep{Capozziello2002,  Santos2007,  DeFelice2010, Campista2011, Capozziello2011}, and higher dimensional braneworld models, in which extra dimension effects drive the current cosmic acceleration by changing the energy balance in a modified Friedmann equation~\citep{Dvali2000, Deffayet2002b, Alcaniz2002, Sahni2003, Maia2005}.
Since very little is known about the nature of the physical mechanism driving the cosmic acceleration, an important way to improve our understanding of this phenomenon is to use cosmological observations to constrain and select its many approaches.

In this paper we use the most recent SNe Ia observations, the joint light-curve analysis (JLA) SNe Ia compilation containing 740 events in the interval $0.01 < z < 1.3$, to perform Bayesian model selection analysis using the {\MN} algorithm~\citep{Feroz2008, Feroz2009, Feroz2013}.
We consider in our analysis different classes of cosmological models and show that a joint analysis involving SNe Ia and BAO data is able to distinguish between the standard cosmology and some of its alternatives.

We organized this paper as follows.
In Sec.~\ref{sec:models} we present the cosmological models considered in our analysis.
The Bayesian framework of model selection is briefly discussed in Sec.~\ref{sec:bayes}.
The data sets and methodology used in the analysis are discussed in Secs.~\ref{sec:data} and~\ref{sec:methodology}, respectively.
We present and discuss the model comparison results in Sec.~\ref{sec:results}.
We summarize our main conclusions in Sec.~\ref{sec:conclusions}.

\section{\label{sec:models}Nonstandard cosmological models}

As mentioned earlier, the late-time cosmic acceleration is usually explored in two different ways: either including an extra component in the right-hand side of Einstein's field equations or modifying gravity at large scales.
In this work, we select models of both cases under the framework of a flat Friedmann-Robertson-Walker (FRW) metric.
In what follows, we briefly discuss the scenarios considered in our analysis.

\subsection{Dark energy models with constant equation of state}

General relativistic scenarios with a constant dark energy EoS $w$ generalize the standard {\lcdm} model in which $w = -1$.
In what follows, we refer to this model as the $w$CDM model.
The corresponding Friedmann equation for this cosmology is given by
\begin{equation}
    \label{constantw}
    E(z)^2 = \OmegaMo a^{-3} + \OmegaDEo a^{-3 (1 + w)} \,,
\end{equation}
where $E(z) = H(z) / H_0$ is the normalized Hubble parameter and $\OmegaMo$ and $\OmegaDEo$ correspond, respectively, to the current values of clustered matter (baryonic and dark) and dark energy density parameters, which obey the normalization condition $\OmegaDEo = 1 - \OmegaMo$.

\subsubsection{Dynamical dark energy models}

A more general case can be studied by allowing the equation of state of the dark energy component to vary as a function of the cosmological scale factor $a$.
In this case, the Friedmann equation takes the form
\begin{equation}
    \label{varyingDE}
    E(z)^2 = \OmegaMo a^{-3} + \OmegaDEo \exp{\left[ 3 \int_a^1 \frac{1 + w(a')}{a'} \, \mathrm{d}a' \right]} \,.
\end{equation}
To discriminate the dynamical dark energy (the time-varying nature of EoS) from that of a cosmological constant $\Lambda$, we consider two kinds of $w(a)$ parametrizations.
First, we consider the Chevallier--Polarski--Linder (CPL) parametrization~\citep{Chevallier2001, Linder2003}, given by
\begin{equation}
    \label{cplwequation}
    w(a) = w_0 + w_a (1 - a) \,,
\end{equation}
where $w_0$ stands for the EoS's value today whereas $w_a$ describes its time evolution.
For this parametrization, the last term of Eq.~(\ref{varyingDE}) is written as
\begin{equation}
    \label{cpl}
    \OmegaDEo a^{-3 (1 + w_0 + w_a)}~\exp{[-3 w_a (1 - a)]} \,.
\end{equation}
As discussed in~\cite{Wang2004}, the above parametrization cannot be extended to the entire history of the Universe since it blows up exponentially in the future ($a \rightarrow \infty$) for $w_a > 0$.
Therefore, we also consider a second dynamical dark energy parametrization suggested by Ref.~\cite{Barboza2008},
\begin{equation}
    \label{bawequation}
    w(z) = w_0 + w_a \frac{z(1 + z)}{1 + z^2} \,,
\end{equation}
which is well behaved over the entire cosmic evolution and mimics a linear-redshift evolution at low redshift.
For this parametrization (referred to it as BA parametrization), the last term of Eq.~(\ref{varyingDE}) can be written as
\begin{equation}
    \OmegaDEo (1 + z)^{3 (1 + w_0)} (1 + z^2)^{\frac{3 w_a}{2}} \,.
\end{equation}
Previous studies have shown that bounds on the $w_0$ and $w_a$ parameters allow this dark energy component to remain subdominant at $z >> 1$.
For details about the classification of different dark energy behaviors using parametrization (\ref{bawequation}), we refer the reader to Ref.~\cite{Barboza2008}.

\subsection{\label{subsec:LtCDM}Vacuum decay model}

An interesting attempt to account for the cosmological constant problems has also been discussed in the context of interacting dark matter and dark energy cosmologies.
A number of ideas have been examined along these lines (see, e.g., \cite{Ozer1986, Carvalho1992, Alcaniz2005, Borges2005, Carneiro2008, Costa2010} and references therein).

The model analyzed in our study has a time-dependent cosmological term $\Lambda(t)$ in which the vacuum energy density $\rho_\Lambda$ decays with the expansion of the Universe as~\citep{Wang2005, Alcaniz2005}
\begin{equation}
    \rho_{\Lambda} = \tilde{\rho}_{\Lambda,0} + \frac{\epsilon\rho_{\mathrm{dm},0}}{3 - \epsilon} a^{-3 + \epsilon} \,,
\end{equation}
where the $\epsilon$ determines the diluting power of the dark matter density $\rho_\mathrm{dm}$ with respect to the usual $a^{-3}$ as $\rho_\mathrm{dm} \propto a^{-3 + \epsilon}$.
Depending upon the positive or negative values of $\epsilon$, the energy is transferred either from dark energy to dark matter or vice versa, respectively.
In such scenarios, dark matter is no longer independently conserved, such that
\begin{equation}
    \label{cont}
    \dot{\rho}_\mathrm{dm} + 3 \frac{\dot{a}}{a} \rho_\mathrm{dm} = - \dot{\rho}_\Lambda \,.
\end{equation}

The Friedmann equation for this class of models is given by~\citep{Ferreira2013}
\begin{equation}
    \label{vacuumdecay}
    E(z) = \left[\OmegaBo a^{-3} + \frac{3 \OmegaDMo}{3 - \epsilon} a^{-3 + \epsilon} + \tilde{\Omega}_{\Lambda,0} \right] \,,
\end{equation}
where $\tilde{\Omega}_{\Lambda,0} = \Omega_{\Lambda,0} - 3 \epsilon \OmegaDMo / (3 - \epsilon)$.
There is an extra degree of freedom compared to the standard {\lcdm} model due to such interaction (for more details on this class of models, see Ref.~\cite{Costa2010}).

\subsection{$f(R)$-gravity models}

The simplest extension of general relativity can be obtained by considering additional terms proportional to powers of the Ricci scalar $R$ in the Einstein--Hilbert Lagrangian, the so-called $f(R)$ gravity.
Differently from general relativistic scenarios, $f(R)$ cosmology can naturally drive an accelerating cosmic expansion without introducing a dark energy field~\citep{Capozziello2002}.

We consider the Einstein--Hilbert action in the Jordan frame including $f(R)$ function of the Ricci scalar as
\begin{equation}
    \label{actionf(R)}
    S = \int \sqrt{-g} \, \frac{f(R)}{2k^2} \, \mathrm{d}^4x + S_\text{matter}(g_{\mu\nu}) \,,
\end{equation}
where $k^2 = 8\pi G$ ($G$ is a bare gravitational constant) and $S_\text{matter}$ represents the action of the matter minimally coupled to gravity.
We assume the metric formalism, in which the connections are assumed to be the Christoffel symbols and the variation of the action is taken with respect to the metric.

In a flat FRW spacetime, the field equations for the action (\ref{actionf(R)}) are given by
\begin{align}
                H^2 &= \frac{k}{3f'} \left( \rho + \frac{Rf' - f}{2} - 3H\dot{R}f'' \right) \,, \\
    2\dot{H} + 3H^2 &= -\frac{k}{f'} \left[ p + \dot{R}^2 f''' + 2H\dot{R}f'' + \ddot{R}f'' +\frac{1}{2} (f - R f') \right] \,,
\end{align}
where a prime denotes derivative with respect to $R$ (we refer the reader to Refs.~\cite{DeFelice2010, Capozziello2011} for more on $f(R)$ cosmologies).
In what follows, we consider the power-law $f(R)$ model \begin{equation}
    \label{frmodel}
    f(R) = R - \beta / R^n \,,
\end{equation}
which satisfies all the viability conditions of $f(R)$ models, as discussed by Ref.~\cite{Pogosian2008}, and reduces to the {\lcdm} model for $n = 0$ and $\beta = 6\OmegaDEo$.

\subsection{DGP model}

The Dvali--Gabadadze--Porrati (DGP) model~\citep{Dvali2000} is an example of an alternative approach which governs cosmic acceleration via modification of Einstein's general relativity, driven by higher dimensional theories.
In this model, our four-dimensional Universe is confined to a three-dimensional brane, embedded in a five-dimensional bulk spacetime with an infinite extra dimension.
The energy-momentum tensor only resides on the brane surface whereas the gravitational field equations are driven by the five-dimensional Einstein tensor and the four-dimensional Einstein tensor of the induced metric on the brane.
Only gravity is allowed to propagate off the 3-brane into the bulk and this induced effect on the brane leads to an accelerated expansion.

A crossover length scale, where the interaction between the effective four-dimensional and five-dimensional gravities takes place, is given by $r_c = M_\mathrm{Pl}^2 / 2 M_5^3$, and the Friedmann equation is modified as~\citep{Deffayet2002b, Alcaniz2002}
\begin{equation}
    \label{Dgp1}
    E(z) = \sqrt{\frac{\rho}{3 M_\mathrm{Pl}^2} + \frac{1}{4 r_c^2}} + \frac{1}{2 r_c} \,,
\end{equation}
where $\rho$ is the energy density of the cosmic fluid.
Note that in the limit of $H \sim r_c^{-1}$, a self-accelerating solution is attained asymptotically, which is the main feature of this model (see Refs.~\cite{Dvali2000, Alcaniz2004} for details).
The above equation can be rewritten as
\begin{equation}
    \label{Dgp2}
    E(z) = \sqrt{\OmegaMo a^{-3} + \Omega_{r_c}} + \sqrt{\Omega_{r_c}} \,.
\end{equation}
Here $\Omega_{r_c}$ represents the density parameter associated with the crossover scale, $\Omega_{r_c} = 1/(4 r_c^2 H_0^2)$.
Under the flat FRW framework, the normalization condition is given by $\Omega_{r_c} = [(1 - \OmegaMo)^2/4]$.
For analysis involving BAO data we add a radiation term, $\Omega_{\gamma,0} = 2.469 \times 10^{-5} h^{-2}$~\citep{Komatsu2011}, to all Friedmann equations above.
A summary of the cosmological models considered in our analysis is given in Table~\ref{tab:models}.

\begin{table}[t]
    \caption{\label{tab:models}Summary of models considered in the analysis along with the free parameters.}
    \begin{ruledtabular}
        \begin{tabular}{lll}
            Model  & Equation                     & Free parameters \\
            \colrule
            \lcdm  & (\ref{constantw}) ($w = -1$) & $\OmegaMo,\, H_0$ \\
            $w$CDM & (\ref{constantw})            & $\OmegaMo,\, H_0,\, w$ \\
            CPL    & (\ref{cplwequation})         & $\OmegaMo,\, H_0,\, w_0,\, w_a$ \\
            BA     & (\ref{bawequation})          & $\OmegaMo,\, H_0,\, w_0,\, w_a$ \\
            \ltcdm & (\ref{vacuumdecay})          & $\OmegaDMo,\, H_0,\, \epsilon $ \\
            $f(R)$ & (\ref{frmodel})              & $\OmegaMo,\, H_0,\, n$ \\
            DGP    & (\ref{Dgp2})                 & $\OmegaMo,\, H_0$ \\
        \end{tabular}
    \end{ruledtabular}
\end{table}

\section{\label{sec:bayes}Bayesian Model Selection}

Bayesian inference is a way to describe the relationship between the model (or hypotheses), the data and the prior information about the model parameters.
In a parameter estimation problem, the starting point for Bayesian data analysis is to compute the joint posterior for a set $\Theta$ of free parameters given the data, $D$, through Bayes' theorem~\citep{Bayes1764}, $P(\Theta | D, M) = \mathcal{L}(D | \Theta, M) \, \mathcal{P}(\Theta | M) / \mathcal{E}(D | M)$, where $P$, $\mathcal{L},$ $\mathcal{P}$ and $\mathcal{E}$ are the shorthands for the posterior, the likelihood, the prior and the evidence,\footnote{Also called Bayesian evidence, marginal likelihood or model likelihood.} respectively.
In short, Bayes' theorem updates our previous knowledge about some model parameters in the light of a given data set.

It is important to note that the evidence $\mathcal{E}$, the denominator of the Bayes' theorem, is just a normalization constant and is uninteresting for parameter estimation, since it does not depend upon the model parameters.
However, in a model comparison problem, the evidence is used to evaluate the model's performance in the light of the data by integrating the product $\mathcal{L}\,\mathcal{P}$ over the full parametric space of the model
\begin{equation}
    \label{eq:evidence}
    \mathcal{E}(D | M) = \int_M \mathcal{L}(D | \Theta, M) \, \mathcal{P}(\Theta | M) \, \mathrm{d}\Theta \,.
\end{equation}
Therefore, the evidence is the average value of the likelihood $\mathcal{L}$ over the entire model parameter space that is allowed before we observe the data.

The most important characteristic of the evidence is its application of Occam's razor to the model selection problem.
It rewards the models that fit the data well and are also predictive, moving the average of the likelihood in Eq.~(\ref{eq:evidence}) towards higher values than in the case of a model which fits poorly or is not very predictive (or is either too complex or has a large number of parameters)~\citep{Liddle2007}.
This concept has been widely applied in cosmology (see, e.g, \cite{Jaffe1996, Hobson2002, Saini2004, Trotta2007, Parkinson2006}).
It is used to discriminate two competing models by taking the ratio
\begin{equation}
    \label{eq:bayesfactor}
    B_{ij} \equiv \dfrac{\mathcal{E}_i}{\mathcal{E}_j} \,,
\end{equation}
which is also known as the Bayes' factor of the model $M_i$ relative to the model $M_j$ (called the reference model in this work).
If each model is assigned equal prior probability, the Bayes factor gives the posterior odds of the two models.

To rank the models of interest, we adopted the scale showed in Table~\ref{tab:jeffreysscale} to interpret the values of $\ln{B_{ij}} = \ln{(\mathcal{E}_i/\mathcal{E}_j)}$ in terms of the strength of the evidence of a chosen reference model.
This scale, suggested by Ref.~\cite{Trotta2008}, is a revised and more conservative version of the Jeffreys scale~\citep{Jeffreys1961}.
Note that the labels attached to the Jeffreys scale are empirical: it depends on the problem being investigated.
Thus, for an experiment for which $|\ln{B_{ij}}| < 1$, the evidence in favor of the model $M_i$ is \emph{usually} interpreted as inconclusive (see Ref.~\cite{Trotta2008} for a more complete discussion about this scale).
Note also that $\ln{B_{ij}} < -1$ means support in favor of the model $M_j$.
In this work, we take {\lcdm} as the reference model $M_j$, so the subscripts $i$ and $j$ are omitted hereafter.

\begin{table}[t]
    \caption{\label{tab:jeffreysscale}The version of the Jeffreys scale for the Bayes factor values discussed in Ref.~\cite{Trotta2008} and adopted in our analysis. The first column shows the threshold values of the logarithm of the Bayes factor of the model $M_i$ relative to the model $M_j$. The second column shows the conventional interpretation of the strength of the evidence above these thresholds.}
    \begin{ruledtabular}
        \begin{tabular}{lc}
            $|\ln{B_{ij}}|$ & Strength of the evidence \\
            \colrule
            $< 1$           & Inconclusive \\
            $1$             & Weak \\
            $2.5$           & Moderate \\
            $5$             & Strong \\
        \end{tabular}
    \end{ruledtabular}
\end{table}

\section{\label{sec:data}Data}

\subsection{Type Ia supernovae}

In this work, we focus primarily on current distance measurements of SNe Ia to perform an observational comparison of the cosmologies discussed in the previous section.
We use the JLA sample which is an extension of the compilation provided by Ref.~\cite{Conley2011} (referred to as the C11 compilation), containing a set of 740 spectroscopically confirmed SNe Ia.
JLA is a compilation of several low-redshift ($z < 0.1$) samples, the full three-year SDSS-II supernova survey~\citep{Sako2014} sample within redshift $0.05 < z < 0.4$, the first three years data of the SNLS survey~\citep{Conley2011, Guy2010} up to redshift $z < 1$ and a few high-redshift Hubble Space Telescope SNe~\citep{Riess2007} in the interval $0.216 < z < 1.755$.
The photometry of SDSS and SNLS was recalibrated and the SALT2 model is retrained using the joint data set.

Theoretically, the distance modulus predicted by the homogeneous and isotropic, flat FRW universe is given by
\begin{equation}
    \label{eq:distance_theory}
    \mu(z, \Theta) = 5\log\frac{d_\mathrm{L}(z, \Theta)}{10\,\mathrm{pc}} \,,
\end{equation}
with the luminosity distance $d_\mathrm{L}$ defined as
\begin{equation}
    \label{eq:ld}
    d_\mathrm{L}(z) = (1 + z) \int_0^z \frac{\mathrm{d}z'}{E(z')} \,,
\end{equation}
where $E(z) = H(z) / H_0$ is the normalized Hubble parameter.
However, from the observational point of view, the distance modulus of a type Ia supernova is obtained by a linear relation from its light curve,
\begin{equation}
    \label{eq:dmobserve}
    \mu = m_B - (M_B - \alpha \times x_1 + \beta \times c) \,,
\end{equation}
where $m_B$ represents the observed peak magnitude in rest-frame $B$ band, $x_1$ is the time stretching of the light curve, and $c$ is the supernova color at maximum brightness.
These three light-curve parameters $m_B$, $x_1$ and $c$ have different values for each supernova and are derived directly from the light curves.
The nuisance parameters $\alpha$ and $\beta$ are assumed to be constants for all the supernovae, but different for different cosmological models.
Following directly Ref.~\cite{Betoule2014}, we also assume a step function relation for the absolute magnitude $M_B$ with the host stellar mass ($M_\text{stellar}$) to compensate the effect of host galaxy properties on $M_B$.
Using Eqs.~(\ref{eq:distance_theory})--(\ref{eq:dmobserve}), one can obtain the predicted magnitude, $m_B(z, \Theta)$, for each one of the cosmological models discussed in the previous section.
The free parameters of our analysis corresponding to the JLA measurements are $\alpha$, $\beta$, $M_B$ and $\Delta_M$.

Using the observed magnitude measurements $m_B(z)$ of the JLA sample (Table F.3 of Ref.~\cite{Betoule2014}) and the predicted ones from Eqs.~(\ref{eq:distance_theory}) and~(\ref{eq:dmobserve}), the Markov chain Monte Carlo (MCMC) simulations for the JLA SNe Ia sample were performed by assuming a multivariate Gaussian likelihood of the type
\begin{equation}
    \mathcal{L}_\text{JLA}(D|\Theta) = \exp[-\chi_\text{JLA}^2(D|\Theta) / 2] \,,
\end{equation}
with
\begin{equation}
    \label{eq:chi2JLA}
    \chi^2_\text{JLA}(\Theta) = \left[ \mathbf{m}_B - \mathbf{m}_B(\Theta) \right]^T C^{-1} \left[ \mathbf{m}_B - \mathbf{m}_B(\Theta) \right] \,,
\end{equation}
where $C$ corresponds to the covariance matrix of the distance modulus $\mu$, estimated accounting for various statistical and systematic uncertainties.
The light-curve fit statistical uncertainties, the systematic uncertainties associated with the calibration, the light-curve model, the bias correction and the mass step uncertainty are described in detail in Sec. 5 of Ref.~\cite{Betoule2014}, whereas the systematic uncertainties related to the peculiar velocity corrections and the contamination of the Hubble diagram by non-Ia are described briefly in Ref.~\cite{Conley2011}.
The uncertainty in redshift due to peculiar velocities, the uncertainty in magnitudes due to gravitational lensing, and the intrinsic deviation in magnitudes are also taken into account while calibrating it.

Using the JLA sample \cite{Betoule2014}, claimed to have provided the most restrictive constraints so far, i.e., $w = -1.027 \pm 0.055$ (assuming $w$ = constant) and $\OmegaMo = 0.295 \pm 0.034$ (for a flat {\lcdm} model).
Therefore, it is interesting to perform a similar analysis for nonstandard cosmological models, calibrating the data to each cosmology and checking their constraining power on the model parameters.

\subsection{\label{subsec:bao}Baryon acoustic oscillations}

Besides the JLA supernovae data set, we also consider in our analysis the measurements of BAO in the galaxy distribution.
The BAO in the primordial plasma have striking effects on the anisotropies of the cosmic microwave background (CMB) and the large scale structure of matter.
The measurements of the characteristic scale of the BAO in the correlation function of matter distribution provide a powerful standard ruler to probe the angular-diameter distance versus redshift relation and the Hubble parameter evolution.
This distance-redshift relation can be obtained from the matter power spectrum and calibrated by the CMB anisotropy data.

Usually, the BAO distance constraints are reported as a combination of the angular scale and the redshift separation.
This combination is obtained by performing a spherical average of the BAO scale measurement and is given by
\begin{equation}
    \label{eq:dz}
    d_z = \frac{r_s(z_\text{drag})}{D_V(z)} \,,
\end{equation}
where
\begin{equation}
    \label{eq:DV}
    D_V(z) = \left[ D_C^2(z) \frac{cz}{H(z)} \right]^{1/3}
\end{equation}
is the volume-averaged distance~\citep{Eisenstein2005} and $D_C(z) = \int_0^z \mathrm{d}z' / H(z')$ is the comoving angular-diameter distance.
In Eq.~(\ref{eq:dz}), $r_s(z_\text{drag})$ is the radius of the comoving sound horizon at the drag epoch $z_\text{drag}$ when photons and baryons decouple~\citep{Eisenstein1998},
\begin{equation}
    \label{eq:rs}
    r_s(z) = \int_{z_\text{drag}}^\infty \frac{c_s(z)}{H(z)} \, \mathrm{d}z \,,
\end{equation}
where $c_s(z) = c / \sqrt{3 \left[ 1 + (3\OmegaBo / 4\Omega_{\gamma,0}) (1 + z)^{-1} \right]}$ is the sound speed in the photon-baryon fluid, and $\OmegaBo = 0.022765 h^{-2}$ and $\Omega_{\gamma,0} = 2.469 \times 10^{-5} h^{-2}$ are the present values of baryon and photon density parameters, respectively, as given by Ref.~\cite{Komatsu2011}.

Table~\ref{tab:data_BAO} shows the BAO distance measurements employed in this work.
In addition to this data, we also include three correlated measurements of $d_z(z=0.44) = 0.073$, $d_z(z=0.6) = 0.0726$ and $d_z(z=0.73) = 0.0592$ from the WiggleZ survey~\citep{Blake2012}, with the following inverse covariance matrix:
\begin{equation}
    C^{-1} =
    \begin{pmatrix}
        1040.3 & -807.5  & 336.8 \\
        -807.5 & 3720.3  & -1551.9 \\
        336.8  & -1551.9 & 2914.9
    \end{pmatrix} \,.
\end{equation}

\begin{table}[t]
    \caption{\label{tab:data_BAO}BAO distance measurements considered in this work.}
    \begin{ruledtabular}
        \begin{tabular}{l d{1.3} d{1.13} c}
            Survey     & \multicolumn{1}{c}{$z$} & \multicolumn{1}{c}{$d_z(z)$} & Reference \\
            \colrule
            6dFGS      & 0.106 & 0.3360 \pm 0.0150 & \cite{Beutler2011} \\
            MGS        & 0.15  & 0.2239 \pm 0.0084 & \cite{Ross2015} \\
            BOSS LOWZ  & 0.32  & 0.1181 \pm 0.0024 & \cite{Anderson2014} \\
            SDSS(R)    & 0.35  & 0.1126 \pm 0.0022 & \cite{Padmanabhan2012} \\
            BOSS CMASS & 0.57  & 0.0726 \pm 0.0007 & \cite{Anderson2014} \\
        \end{tabular}
    \end{ruledtabular}
\end{table}

Using the same methodology applied to the JLA SNe Ia compilation, we also consider a multivariate Gaussian likelihood for the BAO data set.
For each survey listed in the first column of the Table~\ref{tab:data_BAO}, the chi square is given by
\begin{equation}
    \label{eq:chi2BAO_table}
    \chi_\text{survey}^2(D|\Theta) = \left[ \dfrac{d_{z,\text{survey}} - d_z(z_\text{survey}, \Theta)}{\sigma_\text{survey}} \right]^2 \,,
\end{equation}
where $d_{z,\text{survey}}$ and $d_z(z_\text{survey}, \Theta)$ are the observed and theoretical $d_z$, respectively, and $\sigma_\text{survey}$ is the error associated with each observed value.
However, for the WiggleZ data the chi square is of the form
\begin{equation}
    \label{eq:chi2BAO_WiggleZ}
    \chi_\text{WiggleZ}^2(D|\Theta) = \left[ \mathbf{d}_{z,i} - \mathbf{d}_z(\Theta) \right]^T C^{-1} \left[ \mathbf{d}_{z,i} - \mathbf{d}_z(\Theta) \right] \,.
\end{equation}
Then, the BAO likelihood is directly obtained by the product of the individual likelihoods as $\mathcal{L}_\text{BAO} = \mathcal{L}_\text{6dFGS} \times \mathcal{L}_\text{MGS} \times \mathcal{L}_\text{LOWZ} \times \mathcal{L}_\text{SDSS(R)} \times \mathcal{L}_\text{CMASS} \times \mathcal{L}_\text{WiggleZ}$.
Similarly, the joint likelihood for the JLA SNe Ia compilation and the BAO data is given by $\mathcal{L}_\text{joint} = \mathcal{L}_\text{JLA} \times \mathcal{L}_\text{BAO}$.

\section{\label{sec:methodology}Methodology}

While the idea of the Bayes' theorem is simple to understand, the computation of the posterior and the evidence can be difficult both analytically, since the necessary integrals cannot be evaluated in closed form, and numerically, meaning that the integrations can be very time consuming when the dimension of the parametric space is large.
To solve this problem, a widely used practice is to sample from the posterior by applying MCMC techniques (we refer the reader to Refs.~\cite{Metropolis1953, MacKay2003, Skilling2004, Feroz2013} for some MCMC algorithms and to Refs.~\cite{Lewis2002, Mukherjee2006} for applications of some of those algorithms in cosmology).

In this work, we applied an algorithm relying on \PMN\footnote{\url{https://johannesbuchner.github.io/PyMultiNest}.}~\citep{Buchner2014}, a Python\footnote{\url{https://www.python.org}.} interface for the nested sampling (NS) algorithm \MN\footnote{\url{https://ccpforge.cse.rl.ac.uk/gf/project/multinest}.}~\cite{Feroz2008, Feroz2009, Feroz2013}.
NS is designed to directly estimate the relation between the likelihood function and the prior mass, thus obtaining the evidence (and its uncertainty) immediately by summation.
It also computes the samples from the posterior distribution as an optional byproduct.
To compute the evidence values we used the most accurate importance nested sampling (INS)~\cite{Feroz2013} instead of the vanilla NS method, requiring an INS global log-evidence tolerance of 0.1 as a convergence criterion.
Moreover, to improve the accuracy in the estimate of the evidence, we have chosen to perform all analysis by working with a set of 1000 live points, instead of the \MN's default value of 400, so that the number of samples for all posterior distributions was of order $\mathcal{O}(10^4)$.

\begin{table}[t]
    \caption{\label{tab:priors}Priors on the free parameters of each model used to compute the model's evidence. Note that $\N{\mu}{\sigma^2}$ denotes a Gaussian prior with mean $\mu$ and variance $\sigma^2$, and $\U{a}{b}$ denotes the normalized uniform prior for which $\mathcal{P}(x | M) = 1 / (b - a)$ for $a \leq x \leq b$ and $\mathcal{P}(x | M) = 0$ otherwise.}
    \begin{ruledtabular}
        \begin{tabular}{lccc}
            Parameter   & Model associated           & Prior              & Reference \\
            \colrule
            $\OmegaMo$  & All except \ltcdm & $\N{0.3}{0.01}$ & \cite{Feldman2003} \\
            $H_0$       & All & $\N{73.24}{3.028}$ & \cite{Riess2016} \\
            $w$         & $w$CDM & $\N{-1.006}{0.002}$ & \cite{Ade2015} \\
            $w_0$       & BA & $\N{-1.11}{0.063}$ & \cite{Barboza2008} \\
                        & CPL & $\N{-1.005}{0.029}$ & \cite{Hazra2015} \\
            $w_a$       & BA & $\N{0.43}{0.578}$ & \cite{Barboza2008} \\
                        & CPL & $\N{-0.48}{0.593}$ & \cite{Hazra2015} \\
            $\OmegaDMo$ & \ltcdm & $\N{0.26}{0.01}$ & \cite{Komatsu2011, Feldman2003} \\
            $\epsilon$  & \ltcdm & $\N{-0.03}{0.001}$ & \cite{Costa2010} \\
            $n$         & $f(R)$ & $\N{0}{2.5 \times 10^{-13}}$ & \cite{Jain2012, Cataneo2014} \\
            $\alpha$    & All & $\U{0.021}{0.261}$ & \cite{Betoule2014} \\
            $\beta$     & All & $\U{1.601}{4.601}$ & \cite{Betoule2014} \\
            $M_B$       & All & $\U{-19.45}{-18.65}$ & \cite{Betoule2014} \\
            $\Delta_M$  & All & $\U{-0.53}{0.39}$ & \cite{Betoule2014} \\
        \end{tabular}
    \end{ruledtabular}
\end{table}

It is worth mentioning that Bayesian inference (both parameter estimation and model selection) depends on the priors $\mathcal{P}(\Theta | M)$ chosen for the free parameters.
This property accounts for each model's predictive power, turning this dependence in a feature, rather than a defect of Bayesian inference.
Although in Bayesian parameter estimation the use of uniform (flat) priors can be reasonable in some cases, this kind of prior can lead to some issues in a model comparison problem.
Uniform priors with different domain ranges change the evidence and can potentially affect the Bayes factor between two competing models if the models have nonshared parameters.
To use well-motivated priors we considered values that reflect our current state of knowledge about the parameters of the models investigated.
These values are shown in Table~\ref{tab:priors}.\footnote{Note that Ref.~\cite{Jain2012} provides a constraint on the value of $\mathrm{d}f/\mathrm{d}R|_{z=0}$, which can be translated in a constraint on $n$ by differentiating Eq.~(\ref{frmodel}) with respect to $R$.}
We applied uniform priors on the parameters related to the JLA data set ($\alpha$, $\beta$, $M_B$ and $\Delta_M$) since they are common to all models, and so the arbitrary multiplicative constant for these priors cancels out in all Bayes factors.
These uniform priors are centered at the best fit values given by the results involving the JLA data set (stat + sys) as displayed in Table X of Ref.~\cite{Betoule2014}, and have ranges arbitrarily chosen to be 20 times larger than the respective standard deviations as given in that table, a conservative choice to encompass the predictions of all models considered in this work.
For the same reason, we adopted the conservative Gaussian priors $\OmegaMo = 0.3 \pm 0.1$ and $\OmegaDMo = 0.26 \pm 0.1$, since we have fixed $\OmegaBo = 0.022765 h^{-2}$ (see Sec.~\ref{subsec:bao}).
These priors are consistent with model-independent estimates from relative peculiar velocity measurements for pairs of galaxies~\cite{Feldman2003}.

\section{\label{sec:results}Results}

\begin{figure}[t]
    \includegraphics[width=\columnwidth]{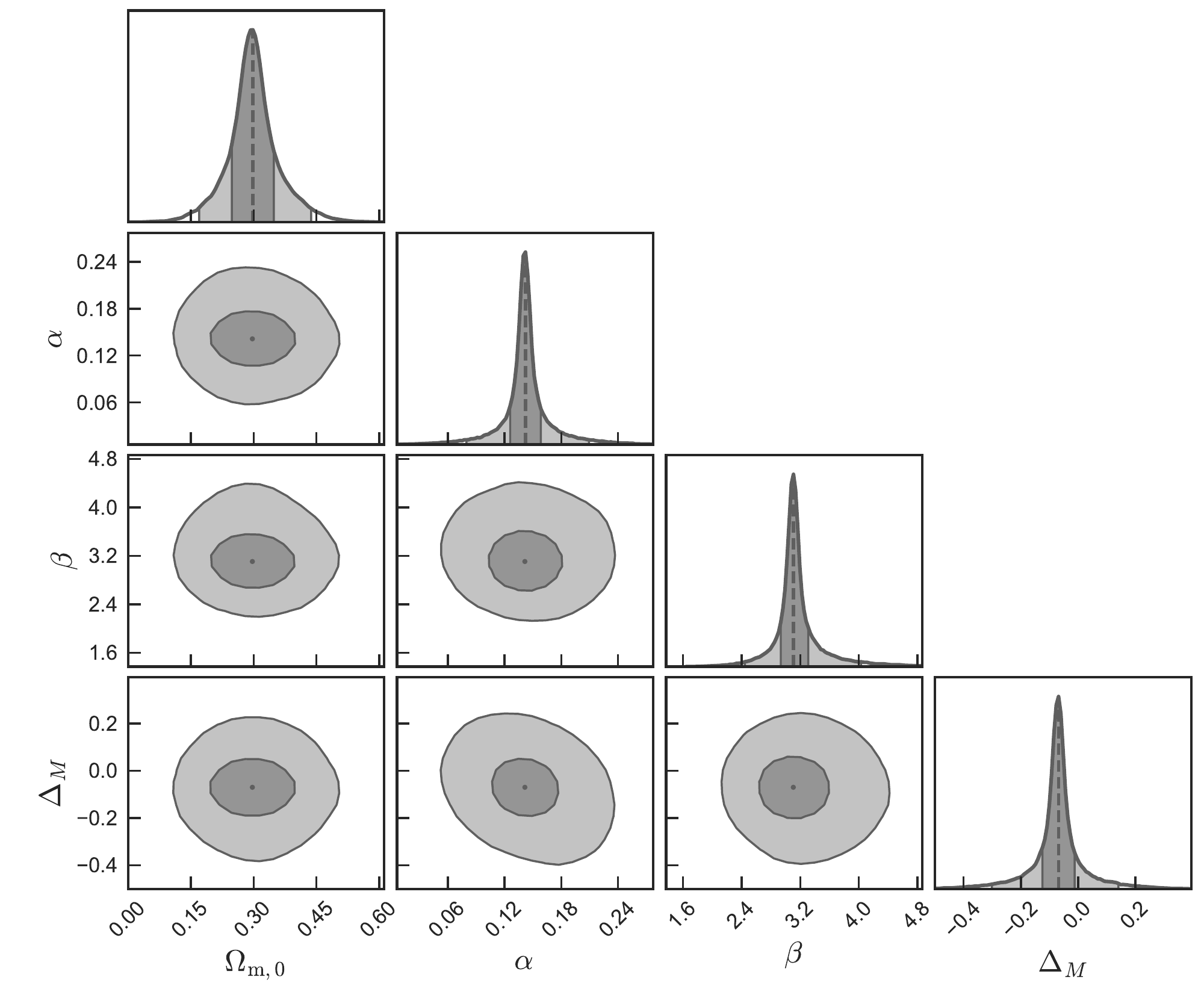}
    \caption{\label{fig:contours_LCDM_JLA}68\% and 95\% credible intervals for the {\lcdm} model using the JLA SNe Ia compilation. The diagonal plots show the posterior distribution for each parameter marginalized with respect to all the other parameters.}
\end{figure}

In Fig.~\ref{fig:contours_LCDM_JLA} we show the parametric space of $\OmegaMo$ and the nuisance parameters $\alpha$, $\beta$ and $\Delta_M$ for the standard {\lcdm} model.
These results were obtained using the JLA SNe Ia sample considering the priors shown in Table~\ref{tab:priors}, as described in the last section.
As shown in the figure, our results are in good agreement with those of Ref.~\cite{Betoule2014} (see Fig.~9 and Table~X of that reference for comparison).
Similar plots for the other cosmological models considered in this analysis are not shown for brevity.

\begin{table}[t]
    \caption{\label{tab:results}Bayesian evidence and Bayes factor for the different cosmologies considered in this work. The results were obtained using the priors shown in Table~\ref{tab:priors}. The last column shows the interpretation of each model's evidence compared to the evidence of the {\lcdm} model, following the Table~\ref{tab:jeffreysscale}.}
    \begin{ruledtabular}
        \begin{tabular}{l d{4.10} d{3.10} c}
            Model  & \multicolumn{1}{c}{$\ln{\mathcal{E}}$} & \multicolumn{1}{c}{$\ln{B}$} & Evidence interpret. \\
            \colrule
            \multicolumn{4}{c}{\rule{0pt}{1.5em}JLA\rule[-1em]{0pt}{0.5em}} \\
            BA     & -357.305 \pm 0.743 &  0.256 \pm 0.743 & Inconclusive \\
            \lcdm  & -357.561 \pm 0.018 &  0               & \dots \\
            $w$CDM & -357.617 \pm 0.027 & -0.056 \pm 0.033 & Inconclusive \\
            \ltcdm & -357.700 \pm 0.014 & -0.139 \pm 0.023 & Inconclusive \\
            DGP    & -357.856 \pm 0.041 & -0.295 \pm 0.045 & Inconclusive \\
            CPL    & -358.034 \pm 0.014 & -0.473 \pm 0.023 & Inconclusive \\
            $f(R)$ & -358.430 \pm 0.041 & -0.869 \pm 0.045 & Inconclusive \\
            \multicolumn{4}{c}{\rule{0pt}{1.5em}BAO\rule[-1em]{0pt}{0.5em}} \\
            CPL    &  -5.837 \pm 0.008 &  0.638 \pm 0.011 & Inconclusive \\
            \ltcdm &  -5.983 \pm 0.008 &  0.492 \pm 0.011 & Inconclusive \\
            $w$CDM &  -6.240 \pm 0.008 &  0.235 \pm 0.011 & Inconclusive \\
            BA     &  -6.292 \pm 0.016 &  0.183 \pm 0.018 & Inconclusive \\
            \lcdm  &  -6.475 \pm 0.008 &  0               & \dots \\
            $f(R)$ &  -7.368 \pm 0.009 & -0.892 \pm 0.012 & Inconclusive \\
            DGP    & -14.981 \pm 0.009 & -8.506 \pm 0.012 & Strong (disfavored) \\
            \multicolumn{4}{c}{\rule{0pt}{1.5em}JLA + BAO\rule[-1em]{0pt}{0.5em}} \\
            \ltcdm & -362.439 \pm 0.023 &   2.945 \pm 0.040 & Moderate \\
            CPL    & -362.878 \pm 0.033 &   2.506 \pm 0.047 & Weak---Moderate \\
            BA     & -363.953 \pm 0.019 &   1.431 \pm 0.038 & Weak \\
            $w$CDM & -364.060 \pm 0.019 &   1.324 \pm 0.038 & Weak \\
            \lcdm  & -365.384 \pm 0.033 &   0               & \dots \\
            $f(R)$ & -365.608 \pm 0.057 &  -0.224 \pm 0.066 & Inconclusive \\
            DGP    & -399.276 \pm 0.075 & -33.892 \pm 0.082 & Strong (disfavored) \\
        \end{tabular}
    \end{ruledtabular}
\end{table}

Our main results are summarized in Table~\ref{tab:results} where the first, second and third subtables correspond to the results obtained using the JLA SNe sample alone, BAO measurements alone and a joint analysis of SNe and BAO, respectively.
These results were obtained considering the priors shown in Table~\ref{tab:priors}.
We first observe that the current SNe Ia data alone cannot rule out any of the cosmological models studied in this analysis.
The joint analysis with BAO data seems to be more effective to this end.
This is clearly seen in the last subtable of Table~\ref{tab:results}, where one can note that, among all, the most dramatic change in the rank of the models with the inclusion of the BAO data in the analysis happens for the DGP model.
Although, as discussed above, one cannot make any conclusions about the evidence of this model in comparison to {\lcdm} from the SNe Ia data alone, the joint analysis with BAO measurements reveals that this scenario is strongly disfavored with respect to the {\lcdm} model.
Using the results from this joint analysis, we see from Eq.~(\ref{eq:bayesfactor}) that, by assuming that the DGP and {\lcdm} models exhaust the model space\footnote{$P(M_i | D) + P(M_j | D) = 1$, where $M_i$ and $M_j$ are the DGP and {\lcdm} models, respectively, and $D$ represents the joint JLA + BAO data set.
Note that $P(M | D) = \int_M \mathcal{P}(\Theta | D, M) \, \mathrm{d}\Theta$.} and keeping their prior probabilities as equal, the probability of the DGP model is not greater than $2.1 \times 10^{-15}$, and the posterior odds in favor of the {\lcdm} model are not less than $\sim 10^{13} : 1$.

\begin{figure}[t]
    \includegraphics[width=\columnwidth]{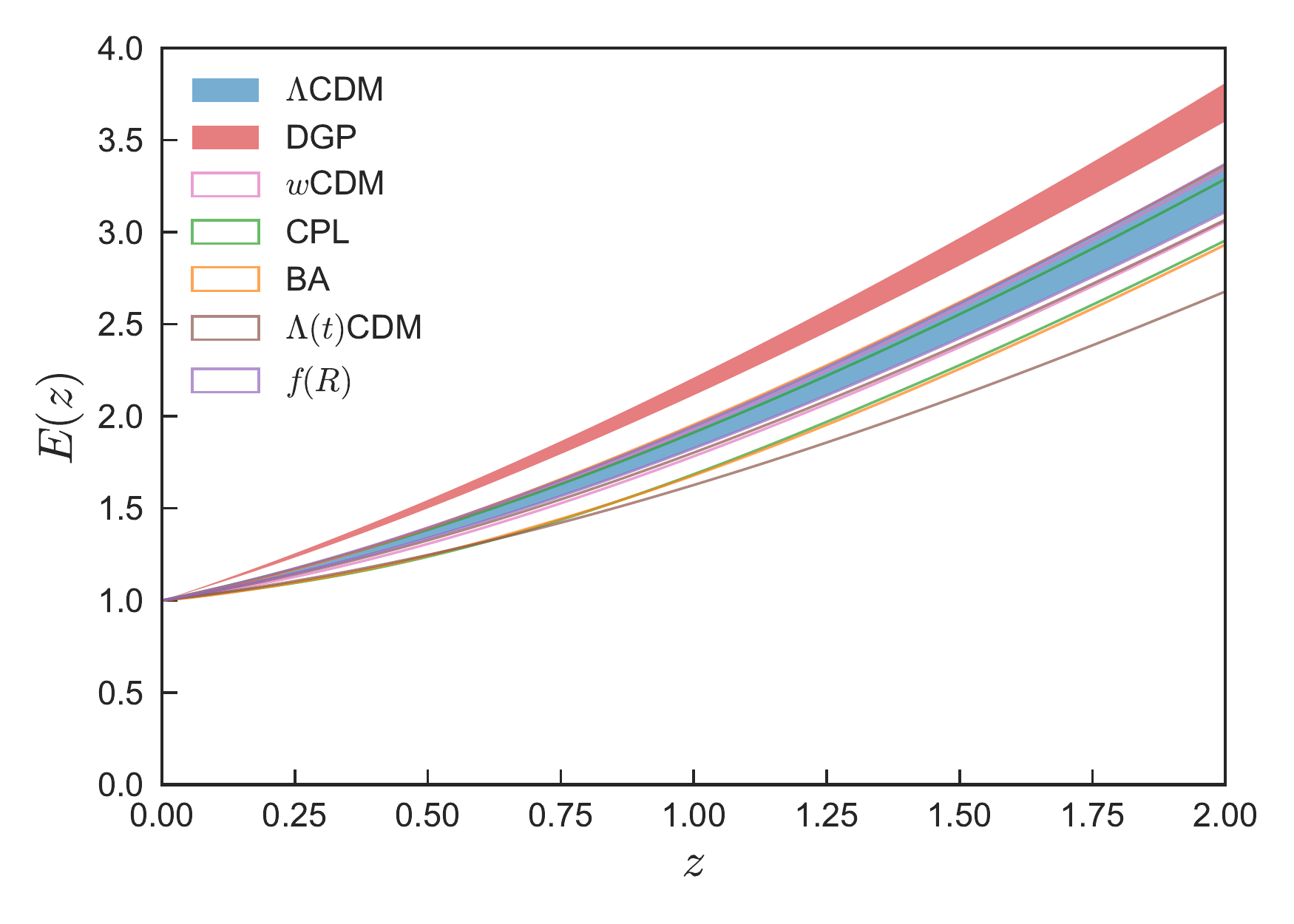}
    \caption{\label{fig:E_evolution}68\% credible intervals for the evolution of $E(z)$ for all the models given in Sec.~\ref{sec:models}.}
\end{figure}

To have some insight into why the DGP model is so significantly disfavored by the data described in Sec.~\ref{sec:data}, we show in Fig.~\ref{fig:E_evolution} the predictive 68\% credible intervals for the evolution of $E(z)$ of all models.
We can see a notable tension involving the evolutions for the DGP model and the evolutions related to the other models, which may explain why this scenario is ruled out in our analyses.

\begin{figure}[t]
    \includegraphics[width=\columnwidth]{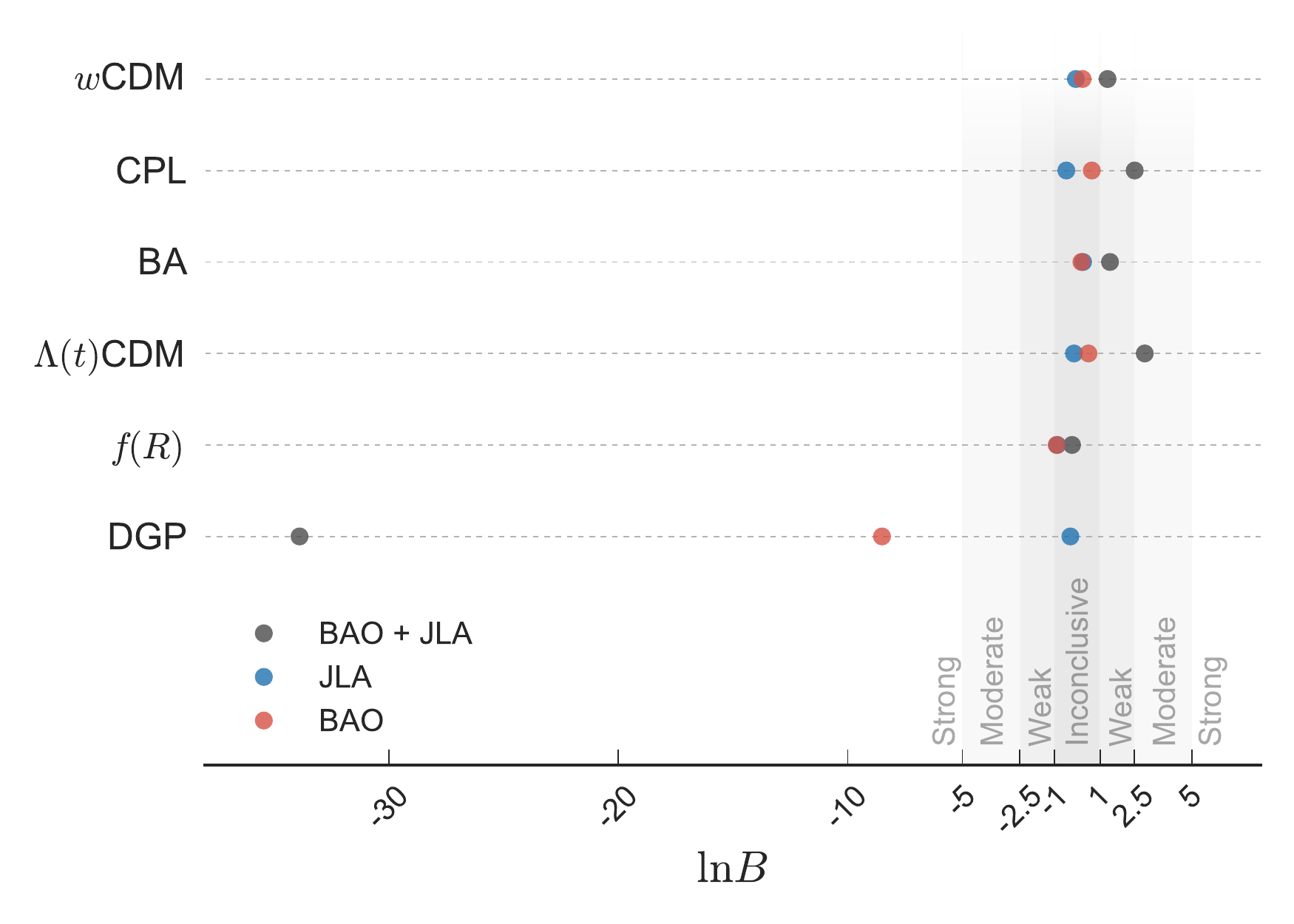}
    \caption{\label{fig:logBFs}Intervals for the Bayes factors between the {\lcdm} and each model given in Table~\ref{tab:models}. The error bars in $\ln{B}$ are not shown since they are much smaller than the marker size for all models. Note that $\ln{B} < -1$ favors the standard {\lcdm} scenario.}
\end{figure}

Regarding the other models, we also note that, with the exception of the Bayes factor related to the DGP model, the joint analysis involving JLA and BAO data shifts all the ranges of the Bayes factors towards a better support for the alternative models compared to {\lcdm}.
One can easily observe these shifts from the graphical representation of the ranges of all Bayes factors, displayed in Fig.~\ref{fig:logBFs}.

The above results should be compared with the ranking order provided by Ref.~\cite{Sollerman2009}.
These authors used 103 SNe Ia from the Sloan Digital Sky Survey-II Supernova Survey along with two data points of the CMB/BAO ratio to rank a number of alternative cosmologies, some of which are also considered in the current study.
Their analysis was performed using two different SNe Ia light-curve fitting, i.e., MLCS and SALT2, and the model ranking was done using the Bayesian and Akaike information criteria.
A comparison of the last subtable of the Table~\ref{tab:results} with the ranking order presented in the Table~II (MLCS) of Ref.~\cite{Sollerman2009} shows a good agreement about the {\lcdm} model and about some of its alternatives as well, but a complete disagreement about the flat DGP case.
In fact, the flat DGP model is at the top in their MLCS ranking.
On the other hand, their Table~III (SALT2) shows an opposite scenario: it rules out the flat DGP model while displaying the flat {\lcdm} model at the top of the rank, unlike our results.
All these differences between our and their results show the influence of the SNe Ia light-curve fitting on the parameter estimation and model selection.

Another similar work was done in Ref.~\cite{Rubin2009}.
The authors compared several nonstandard cosmological models by performing a maximum likelihood analysis combining 307 SNe Ia from the Union08 compilation with constraints from BAO and CMB measurements.
Although sharing only three models with our work (namely, the {\lcdm}, $w$CDM and DGP cosmologies), the ranking order displayed in Table~I of Ref.~\cite{Rubin2009} seems to be more consistent with our results, showing that the $w$CDM model is better ranked than the standard {\lcdm} scenario, while the DGP alternative performs worse compared to all other models studied in their analyses.

\section{\label{sec:conclusions}Conclusions}

Given the current state of uncertainty that remains over the physical mechanism behind the observed acceleration of the Universe, an important way to improve our understanding of this phenomenon is to use cosmological observations to constrain its different approaches.
In this paper, we have performed a Bayesian model selection statistics to rank some nonstandard cosmological models in the light of the most recent SNe Ia (JLA compilation) and BAO data.
Our analyses have shown that the JLA data alone are unable to distinguish between the standard {\lcdm} scenario from some specific examples of coupled quintessence cosmologies [\ltcdm], modified gravity models [$f(R)$ and DGP] and simple parametrizations of the dark energy component.
On the other hand, while not being able to distinguish most of the alternative models considered in this work, the current BAO measurements can strongly rule out the flat DGP model (see Table~\ref{tab:results}).

We have also shown that, when a joint analysis involving SNe Ia and BAO data is performed, the evidence for the DGP model is weakened with respect to the {\lcdm} model.
The result of this joint analysis shows that the DGP scenario becomes even more strongly disfavored with respect to the standard cosmology, with $\ln{B} = -33.892 \pm 0.082$, whereas the analysis using the BAO data alone provides $\ln{B} = -8.506 \pm 0.012$ and $\ln{B} = -0.295 \pm 0.045$ from the JLA data alone.
These results are consistent with some of the previous studies done using different statistics and data sets (see, e.g., Refs.~\cite{Sollerman2009, Rubin2009}).

Finally, an important aspect worth emphasizing concerns the ranking position of the decaying vacuum cosmology considered in our analysis.
As discussed earlier (see Sec.~\ref{subsec:LtCDM}), in this kind of model the dark energy field interacts with the pressureless component of dark matter in a process that violates adiabaticity and that constitutes a phenomenological attempt at alleviating the coincidence problem~\citep{Wang2005, Alcaniz2005}.
We have found that this scenario provides an excellent fit to both SNe Ia observations and SNe Ia plus current baryon acoustic oscillation measurements.

\begin{acknowledgments}
    B.~S. and N.~C.~D. are supported by the National Observatory DTI-PCI program of the Brazilian Ministry of Science, Technology and Innovation (MCTI).
    N.~C.~D. also acknowledges support from the Dirección General de Asuntos del Personal Académico, Universidad Nacional Autónoma de México (DGAPA-UNAM) postdoctoral fellowship.
    J.~S.~A. thanks Conselho Nacional de Desenvolvimento Científico e Tecnológico (CNPq), Fundação Carlos Chagas Filho de Amparo à Pesquisa do Estado do Rio de Janeiro (FAPERJ) and Coordenação de Aperfeiçoamento de Pessoal de Nível Superior (INEspaço/CAPES) for the financial support.
\end{acknowledgments}

\bibliography{ref}

\end{document}